\documentclass[twocolumn]{aastex631}

\usepackage{animate}
\usepackage{subfigure}
\usepackage{amsmath}
\usepackage{amssymb}
\usepackage{comment}
\usepackage[T1]{fontenc}
\usepackage{graphicx}% Include figure files
\usepackage{booktabs}
\usepackage{dcolumn}% Align table columns on decimal point
\usepackage{bm}% bold math
\usepackage{hyperref}% add hypertext capabilities
\usepackage{ragged2e}
\usepackage{academicons}
\usepackage{xcolor}
\usepackage{physics}
\hypersetup{colorlinks=True, linkcolor=blue, filecolor=blue, urlcolor=blue, citecolor=blue}
\newcommand{\cmmnt}[1]{\ignorespaces}
\newcommand{\vect}[1]{\boldsymbol{#1}}

\def\mel{m_{{\rm e}}}
\def\mp{m_{\rm p}}
\def\nel{n_{{\rm e}^-}}
\def\np{n_{\rm p}}
\def\Thetae{\Theta_{\rm e}}

\def\Lj{L_{\rm j}}
\def\Rj{r_{\rm j}}
\def\rhoj{\rho_{\rm j}}
\def\vj{v_{\rm j}}

\def\rhoc{\rho_{\rm c}}
\def\Rc{R_{\rm c}}
\def\el{{\rm e}^-}
\def\pos{{\rm e}^+}
\def\pr{{\rm p}^+}

\usepackage{ulem}

\shorttitle{MHD 3D simulations of FR I Jets}
\shortauthors{Tripathi et al.}
\begin{document}

\title{How Plasma Properties of the Fanaroff-Riley Jet can Shape its Morphology}
\correspondingauthor{Indranil Chattopadhyay}
\email{indra@aries.res.in}

\author[0009-0002-7498-6899]{Priyesh Kumar Tripathi}
 \affiliation{Aryabhatta Research Institute of Observational Sciences (ARIES), Manora Peak, Nainital, 263001, India\\}
 \affiliation{Department of Applied Physics, Mahatma Jyotiba Phule Rohilkhand University, Bareilly, 243006, India\\}

\author[0000-0002-2133-9324]{Indranil Chattopadhyay}
 \affiliation{Aryabhatta Research Institute of Observational Sciences (ARIES), Manora Peak, Nainital, 263001, India\\}

\author[0000-0002-9036-681X]{Raj Kishor Joshi}
 \affiliation{Nicolaus Copernicus Astronomical Center, Polish Academy of Sciences, Bartycka 18, PL-00-716 Warsaw, Poland\\}

\author[0000-0001-9899-7686]{Ritaban Chatterjee}
 \affiliation{School of Astrophysics, Presidency University, 86/1 College Street, Kolkata, West Bengal, India 700073\\}
 
\author[0000-0002-9851-8064]{Sanjit Debnath}
 \affiliation{Aryabhatta Research Institute of Observational Sciences (ARIES), Manora Peak, Nainital, 263001, India\\}

\author{M. Saleem Khan}
 \affiliation{Department of Applied Physics, Mahatma Jyotiba Phule Rohilkhand University, Bareilly, 243006, India\\}


\begin{abstract}
   Extragalactic jets are broadly classified into two categories based on radio observations: core-brightened jets, known as Fanaroff-Riley Type I (FR I), and edge-brightened jets, classified as Type II (FR II).  
   This FR dichotomy may arise due to variation in the ambient medium and/or the properties of the jet itself, such as injection speed, temperature, composition, magnetization, etc.
   To investigate this, we perform large-scale three-dimensional magnetohydrodynamic (3D-MHD) simulations of low-power, supersonic jets extending to kiloparsec scales. We inject a jet beam carrying an initially toroidal magnetic field into a denser, unmagnetized, and stratified ambient medium through a cylindrical nozzle. Our simulations explore jets with varying injection parameters to investigate their impact on morphology and emission properties. Furthermore, we examine jets with significantly different plasma compositions, such as hadronic and mixed electron-positron-proton configurations, to study the conditions that may drive transitions between FR I and FR II morphologies. We find that, under the same injection parameters, mixed plasma composition jets tend to evolve into FR I structures. In contrast, electron-proton jets exhibit a transition between FR I and FR II morphologies at different stages of their evolution.
\end{abstract}


%\keywords{\uat{Galaxies}{573} --- \uat{Cosmology}{343} --- \uat{High Energy astrophysics}{739} }


%%%%%%%%%%%%%%%%%%%%%%%%%%%%%%%%%%%%%%%%%%%%%%%%%%

%%%%%%%%%%%%%%%%% BODY OF PAPER %%%%%%%%%%%%%%%%%%

\section{\label{sec:Introduction}Introduction}
   The extragalactic jets powered by supermassive black holes (SMBHs) at the centers of active galactic nuclei (AGNs) emit radiation across the entire range of the electromagnetic spectrum. The emission produced from these jets spans from low-energy radio bands to extremely high-energy gamma-ray bands, making them some of the most luminous sources in the universe \citep{2019_Blandford_et_al_ARA&A..57..467B}.
   Based on the radio observations, \citet{1974_Fanaroff_Riley_MNRAS} broadly classified these astrophysical jets into two categories: Fanaroff-Riley Type I (FR I) and Type II (FR II). FR I jets display core-brightened diffuse structures, which become fainter as one approaches the outer lobes, whereas FR II jets are characterized by powerful, edge-brightened giant lobes with strong hotspots, where most of the emission is localized (for a review, see \citet{1984_Bridle_Perley_ARA&A..22..319B}). 
   Some of the possible explanations for this widely debated FR I/II dichotomy are: (i) interaction of the jet plasma with that of the ambient medium could cause a transition of an initially supersonic jet to a subsonic flow thereby decelerating it \citep{1993_DeYoung_ApJ,1994_Komissarov_MNRAS,1995_Bicknell_ApJS}, or (ii) due to differences in the central engine itself (spin of the BH or the accretion mode) \citep{1992_Baum_ApJ,1995_Baum_ApJ,1999_Meier_ApJ,2007_Wold_A&A}. Another possible explanation could be the difference in the underlying plasma composition of jets \citep{2018_Croston_MNRAS}, which may vary from pure pair-plasma (composed of electrons and positrons) to a mixed composition involving electrons, positrons, and protons or even jets consisting of electrons and protons only. The pair plasma ($\el~-~\pos$) may be favored for FR I sources \citep[as suggested for M87 jet][]{1996_Reynolds_MNRAS}, while electron-proton ($\el~-~\pr$) jets may be preferred for FR II sources \citep{1993_Celotti_MNRAS,1997_Celotti_MNRAS}.
   
   Interestingly, observations show that both FR I and FR II jets are highly relativistic at the parsec (pc) scale, having nearly similar jet velocities \citep{2001_Giovannini_et_al_ApJ,2008_Celotti_Ghisellini_MNRAS}. While on kiloparsec (kpc) scales, FR II jets maintain relativistic velocities, whereas FR I jets undergo significant deceleration, transitioning to subrelativistic and subsonic speeds and thus terminate gently without forming hotspots \citep{2023_GopalKrishna_JApA}. This indicates that the nature of the central engine is the same for both FR I and FR II sources. Moreover, the deceleration in FR I jets must occur within the intermediate region between the jet base and the kpc scale \citep{1999_Laing_et_al_MNRAS,2002_Laing_Bridle_MNRAS,2014_Laing_Bridle_MNRAS,2016_Tchekhovskoy_Bromberg_MNRAS}.
   Additionally, the existence of HYMORS (HYbrid MOrphology Radio Sources) \citep{2000_Gopal-Krishna_WiitaA&A,2002_Gopal-Krishna_Wiita_NewAR,2023_GopalKrishna_JApA}, whose display different radio morphology on the two sides of the central engine (one radio lobe appears FR I while the other appears FR II), supports explanations for the FR dichotomy based on jet interaction with the external medium rather than the central engine.
   
   Several numerical works have been done (see review by \citet{2021_Komissarov_Porth_NewAR}) to address this FR dichotomy related to both deceleration and propagation \citep{2010_Perucho_et_al,2007_Perucho_Marti,2009_Meliani_Keppens,2014_Perucho_et_al,2017_Toma_et_al,2018_Gourgouliatos_Komissarov,2016_Tchekhovskoy_Bromberg_MNRAS,2020_Mukherjee_et_al}. A detailed study on the propagation of already decelerated low Mach number jets was carried out by \citet{2016_Massaglia_et_al_A&A,2019_Massaglia_et_al_A&A,2022_Massaglia_et_al_A&A}, successfully simulating jet structures similar to those in FR I radio sources. They also showed the necessity of three-dimensional simulations to capture the transition to diffuse and turbulent flow structures. \citet{2020_Rossi_et_al_A&A,2024_Rossi_et_al} further investigated jet deceleration within the galaxy core (sub-kpc scale) under varying jet-ambient density ratios and found the jet injection parameters of \citet{2016_Massaglia_et_al_A&A} to be generally consistent. An important limitation of these simulations is the lack of detailed information about the matter content of the jet. So far, only a few studies have explored this aspect. \citet{scheck02} presented the first long-term 2D relativistic simulations of multi-species jets, including both purely leptonic and baryonic plasmas. While their resolution was sufficient to capture the overall morphological structure, it was probably inadequate due to the assumption of 2D. 
   Recently, \cite{2023_Joshi_Chattopadhyay_ApJ...948...13J} performed the 2D RHD simulations of jets at moderate resolutions (10 cells to resolve the jet radius at the injection nozzle). These simulations covered a wide range of parameter space with different initial conditions to investigate the effect of plasma composition on large-scale morphology and showed that jets launched with different compositions do not evolve identically despite fixing the same macroscopic injection parameters. While the results exhibited quantitative differences, all of the models evolved as FR II type jets —likely a consequence of the simulations being two-dimensional, which prevents the development of turbulent cascades to smaller scales. Moreover, the magnetic field was ignored in this work. 

   In the present paper, we set the jet base at the scale of kpc distance from the central SMBH, where the jet has already been decelerated to subrelativistic speeds by some means, and therefore, we chose to use non-relativistic MHD framework to study further evolution of the jet. A non-relativistic jet automatically does not become an FR I jet. The aim of the present work is to investigate the influence of various jet plasma properties, including plasma composition, that may influence the transition from FR II to FR I jet.

   The outline of the paper is as follows. Section~\ref{sec:Simulation Setup} describes the governing equations of our simulations, the use of the relativistically correct equation of state, the description of the simulation code, and the initial setup of our simulations. Then, we present our results in Section~\ref{sec:Results}, followed by the summary and discussions in Section~\ref{sec:Summary and Discussions}.

%%%%%%%%%%%%%%%%%%%%%%%%%%%%%%%%%%%%%%%%%%%%%%%%%%
\section{\label{sec:Simulation Setup}Simulation Setup}
   In this section, we describe the setup of our MHD simulations, i.e., equations of motion and details of the code used.
\subsection{\label{sec:Equations of magnetohydrodynamics}Equations of magnetohydrodynamics}
   We perform numerical simulations of the non-relativistic jets, which are described by the time-dependent ideal MHD equations of motion. The coordinate-free conservative form of these equations for inviscid flow is:
      \begin{subequations}
      \label{eq:MHD eqns}
        \begin{align}
          & \frac{\partial \rho}{\partial t} + 
                            \vect{\nabla} \cdot (\rho \vect{v}) = 0 
      \\  & \frac{\partial (\rho \vect{v})}{\partial t} + 
                       \vect{\nabla} \cdot (\rho\vect{v \otimes v} - \vect{B \otimes B} + p^*\vect{I}) = 0
      \\  & \frac{\partial E}{\partial t} + 
                 \vect{\nabla} \cdot \left[(E+p^*)\vect{v} - \vect{B(v \cdotp B)}\right] = 0 
      \\  & \frac{\partial \vect{B}}{\partial t} + 
                 \vect{\nabla} \cdot (\vect{v \otimes B - B \otimes v}) = 0 
      \\  & \frac{\partial (\rho \Phi)}{\partial t} + 
                 \vect{\nabla} \cdot (\rho \vect{v} \Phi) = 0 ,
        \end{align}
      \end{subequations} 
      
  where $\rho, \vect{v}$, and $\vect{B}$ denote the mass density, bulk velocity, and magnetic field, respectively, $\vect{I}$ is a unit tensor, $p^* = p_{th} + B^2/2$ is total pressure (sum of thermal pressure and magnetic pressure), and $E = e + (\rho v^2 + B^2)/2$ is the total energy density or the sum of thermal energy density, the kinetic energy density, and the magnetic energy density of the fluid element. Additionally, an equation for a tracer field (denoted as $\Phi$) is solved to distinguish between the jet material and ambient medium. The transport of the passive tracer field obeys the simple advection equation of the form,
       \begin{equation}
          \frac{\partial \Phi}{\partial t} + 
                            (\vect{v \cdotp \nabla}) \Phi  = 0 
        \label{eq:tracer_field}
       \end{equation}

\subsection{\label{sec:Equation of state}Equation of state (EoS)}
  Further, to close the system of equations \eqref{eq:MHD eqns}, we additionally need an equation of state. We use an approximate yet relativistically correct EoS, given by \citet{2009Chattopadhyay&RyuApJ} (CR EoS) with variable adiabatic index ($\Gamma$) for multispecies flows, which is a close fit to the first relativistically correct EoS given by \citet{Chandrasekhar1939}. To account for the transition between very high to very low temperatures in the flow, the EoS should carry the temperature information of the flow. In addition, the composition is also important for a flow to be thermally relativistic or non-relativistic, depending on whether the thermal energy $(kT)$ is comparable to its rest mass energy $(mc^2)$. The CR EoS determines the thermodynamics from both the temperature and composition of the flow. The internal energy in the CR EoS in the unit system $c=1$ is given as $e = \rho f $ where, 
       \begin{equation}
          f = 1 + (2-\xi)\Theta \left[\frac{9\Theta+6/\tau}{6\Theta+8/\tau}\right] + 
            \xi \Theta \left[\frac{9\Theta+6/\eta \tau}{6\Theta+8/\eta \tau}\right]
        \label{eq:CR EoS}
       \end{equation}
     In the above equation, $\xi = \np/\nel$ and $\eta = \mel/\mp$ are ratios of the number density of the proton to electron and mass of the electron to the proton, respectively. In addition, $\tau = 2-\xi + \xi /\eta$ , which is a function of the composition ($\xi$) of the flow. $\Theta = p/\rho$ is a measure of temperature (in physical units, it is the ratio of thermal energy with the rest mass energy or $2kT/ (\mel \tau c^2)$). The specific enthalpy is given as $h = (e + p)/\rho = f + \Theta$ and the polytropic index $N$ is calculated as, $N = \rho (\partial{h} / \partial{p}) - 1  = \partial{f} / \partial{\Theta} $. The adiabatic index will be, $\Gamma = 1 + {1}/{N}$. For a more detailed description of the equations, one may refer to \citet{2021joshietal_MNRAS.502.5227J}. CR EoS has been implemented by many authors to study various scenarios \citep{2011_Chattopadhyay_Chakrabarti_IJMPD..20.1597C,2014_Cielo_etal_MNRAS.439.2903C,2019SinghChattopadhyay,2019SinghChattopadhyay2,2021joshietal_MNRAS.502.5227J,2022_Joshi_etal_ApJ...933...75J,2023_Joshi_Chattopadhyay_ApJ...948...13J,2024_Debnath_et_al_MNRAS.528.3964D,2025Debnath_et_al,2024_Joshi_et_al_ApJ...971...13J,2025Tripathi_et_al_979_61}. The present form of EoS assumes that the positive and negative charges of the plasma settles into same temperature distribution. Similar approximation is also assumed for the polytropic EoS (i.e., $e=\rho \Theta/(\Gamma -1)$), which is used in most of the jet simulation papers cited here. The two-temperature version of CR EoS has also been presented in several papers \citep{2020Sarkar_etal,2023_Sarkar_etal_MNRAS.522.3735S}. A fully ionized plasma is composed of ions and electrons, i.e., composed of dissimilar particles, and the radiative properties of ions and electrons are significantly different. If electrons and ions relax into a single temperature distribution, then the electron-proton interaction time scale ($\mathcal{T}_{\rm ep}$) should be far shorter than cooling/heating timescales and most importantly the advection time scale ($\mathcal{T}_{\rm adv}\sim z/v_z$, where $z$ and $v_z$ represent the characteristic length scale and velocity, respectively). It is well known that astrophysical plasma seldom shows $\mathcal{T}_{\rm ep} < \mathcal{T}_{\rm adv}$. 
     Therefore, it is fair to say that the single temperature description for fully ionized plasma is essentially an approximation.

\subsection{\label{sec:Code description}Code description}
    We utilize our recently developed multi-dimensional simulation code for this work, which solves the ideal magnetohydrodynamic equations \eqref{eq:MHD eqns} using finite volume methods. It employs second-order Godunov-type schemes to piecewise linearly (PLM) reconstruct the cell-centered primitive variables with a min-mod limiter. The fluxes are calculated using HLLD approximate Riemann solver \citep{2005_Miyosh_Kusano_JCP}, and finally, the second-order total variation diminishing Runge-Kutta (TVD-RK2) scheme is used for the time integration. In addition, the hyperbolic divergence cleaning scheme \citep{2002Dedner_etal_JCoPh.175..645D} is employed to maintain the solenoidal constraint on the magnetic field ($\vect{\nabla} \cdotp \vect{B} = 0$). The parallel processing capability of the code is implemented through the method of domain decomposition using the Message Passing Interface (\texttt{MPI}) library.
    
       \begin{ruledtabular}
      \begin{table*}
	      %\centering
      	\caption{Details of the parameters used in the simulations}
	      \label{tab:siml_param}
	      \begin{tabular}{lccccccccc}
	      	Model & $\rhoj/\rhoc$ & $T_c (keV)$ & $\vj (c)$ & $\beta$ & $\xi$ & $\mathcal{M}_{\rm j}$ & $L_{\rm j}$ (erg/sec)  & Notes & Morphology \\ 
	      	\tableline
	      	REF  & $10^{-3}$ & 0.167  & 0.067 & 10  & 1.0  & 4 & $2.01 \times 10^{42}$ & Reference case & FR I/II \\
                HYP  & $10^{-3}$ & 0.167   & 0.167 & 10  & 1.0  & 10 & $3.14 \times 10^{43}$ & highly supersonic & FR II \\
                HOT  & $10^{-3}$ & 0.305   & 0.067 & 10  & 1.0  & 3 & $2.01 \times 10^{42}$ & same velocity, hotter medium & FR I \\
                MAG  & $10^{-3}$ & 0.167  & 0.067 & 4  & 1.0  & 4 & $2.01 \times 10^{42}$ & high magnetization & FR I/II \\
                CMp5  & $10^{-3}$ & 0.167   & 0.067 & 10  & 0.5 & 2.87  & $2.01 \times 10^{42}$ & mixture of electron, proton and positron & FR I \\
                CMp2  & $10^{-3}$ & 0.167   & 0.067 & 10  & 0.2 & 1.83  & $2.01 \times 10^{42}$ & " & FR I \\
%                CM0  & $10^{-3}$ & 0.167  & 0.14  & 0.067 & 10  & 0.0   & $0.00 \times 10^{42}$ & pair plasma & FR I? \\
	      \end{tabular}
%            \tablenotemark{key letter(s)}
            \tablenotetext{}{\textbf{Note.} \justifying Here $T_c$ represents the core temperature of the ambient medium. The variables $\vj (c)$, $\beta$, and $\mathcal{M}_{\rm j}$ represent the velocity, plasma-$\beta$ parameter, and the Mach number of injected jet flow, respectively. $\xi$, $L_{\rm j}$ represent the composition parameter of the flow, and jet kinetic power, respectively.}
      \end{table*}       
   \end{ruledtabular}

\subsection{Initial setup and Boundary conditions}
   In our simulations, we chose the jet beam radius ($\Rj$) as the unit of length, the speed of light ($c$) as the unit of velocity, and the jet density ($\rhoj$) as the unit of density. So the unit of time is $\Rj/c$. We conducted our 3D simulations in a cartesian computational box of size $x \in [-20,20]$, $y \in [-20,20]$, and $z \in [0,160]$ in units of the jet beam radius $\Rj$. The computational box was covered with $240 \times 240 \times 960$ uniformly spaced cells, giving a resolution of 6 cells per beam radius. The computational domain was initially filled with an unmagnetized medium at rest with uniform gas pressure $p_{\rm a}$. The density of the medium was spherically stratified according to a King-like profile \citep{1972_King_ApJ} as:
       \begin{equation}
          \rho(R) = \frac{\rhoc}{\rhoj} \frac{1}{1+(R/\Rc)^2} 
        \label{eq:king_profile}
       \end{equation}
   where $\rhoj$ and $\rhoc$ are the jet density and the core density, $R=\sqrt{x^2+y^2+z^2}$ is the spherical radius, $\Rc$ is the core radius which we set to 40$\Rj$. 
   A jet beam was continuously injected into the ambient medium through a circular base of unit radius $(r = \sqrt{x^2+y^2} < 1)$ at the lower z-boundary $(z=0)$. The inflow jet beam has constant density $\rhoj$, a velocity component in the z-direction $\vj$, and the tracer field $\Phi_{\rm j}$ set to unity. The injected jet beam also carries a purely toroidal (azimuthal) magnetic field as given in \citet{1989_Lind_et_al_ApJ...344...89L}:
      \begin{eqnarray}
          B_{\phi} = \begin{cases}
                      -B_{\rm m} r/a, & \text{if $r < a$};\\
                      -B_{\rm m} a/r, & \text{if $r > a$}.
                     \end{cases}
       \label{eq:magnetic_field_profile}
      \end{eqnarray}
   where $a$ is the magnetization radius, which is chosen to be 0.5,
   and $B_{\rm m}$ is the maximum magnetic field strength at $r = a=0.5$. This magnetic field configuration corresponds to a constant current inside $r=a$ and zero outside. The jet base was kept in pressure equilibrium with the ambient medium; therefore, the pressure profile of the jet base was determined by solving the radial component of the momentum balance equation. At the jet base, $v_r = v_{\phi}=0$ and $B_r = B_z=0$, the radial momentum balance equation reduces to, 
       \begin{equation}
           \frac{dp}{dr} =  -\frac{d}{dr} \left(\frac{B_{\phi}^2}{2}\right) - \frac{B_{\phi}^2}{r}
           \label{eq:diff_pressure}
       \end{equation}
   Now, piecewise solving equation~\eqref{eq:diff_pressure} by taking the magnetic field profile given by equation~\eqref{eq:magnetic_field_profile}, we get the pressure equilibrium condition as,
       \begin{equation}
          p(r)|_{z=0} = p_{\rm a} + B_{\rm m}^2 \left[1-\text{min}(r^2/a^2,1)\right]
        \label{eq:jet_pressure_profile}
       \end{equation}
   So equation~\eqref{eq:jet_pressure_profile} is the radial pressure profile of the injected plasma at the jet base.
   The reflective boundary conditions were imposed outside the nozzle to mimic a counterjet. The other boundaries in the simulation were kept as zero-gradient outflow boundaries, where the matter can leave the domain.
   
   We conducted a set of six simulations: one reference case and then changing different parameters. The details of the parameters are listed in Table~\ref{tab:siml_param}. We calculate the jet kinetic power in the limit of subrelativistic and matter-dominated jets (plasma-$\beta > 1$), which can be defined as:
       \begin{equation}
          \Lj = 0.5 \rhoj \vj^2 ~ \vj ~ \pi \Rj^2 
        \label{eq:jet_power}
       \end{equation}
   where $\rhoj$, $\vj$, and $\Rj$ are the density, velocity, and beam radius of the jet.
   
%%%%%%%%%%%%%%%%%%%%%%%%%%%%%%%%%%%%%%%%%%%%%%%%%%
\section{\label{sec:Results}Results}
    Our aim is to investigate whether the matter content of extragalactic jets can influence their morphological characteristics. 
    For this purpose, we simulate low-power jets that have already decelerated to subrelativistic speeds.
    Assuming the jet beam radius ($\Rj$) to be 100 pc, the galactic core radius ($\Rc$) in our simulations will be at 4 kpc from the jet base. As shown by X-ray observations, the average temperature within the galactic core is typically $T_c \sim 0.1-0.3~keV$ \citep{2013_Posacki_et_al}. The unit time in our code corresponds to $\sim 326$ years. 
    We begin with simulating a reference case (Model-REF), which has $T_c \sim 0.167~keV$, followed by a series of simulations exploring the effects of varying key parameters: injection velocity (Model-HYP), the temperature of the jet and the medium (Model-HOT), and the jet magnetization (Model-MAG). Finally, we explore the impact of different jet plasma compositions in Model-CMp5 and CMp2.

    Further, to check the emission properties in our models, we compute the posterior synchrotron emission from our simulation results. We calculate the thermal synchrotron emissivity distribution (in erg cm$^{-3}$ s$^{-1}$) as described in \citet{1983_Shapiro_Teukolsky_bhwd.book.....S} as,
      \begin{equation}
          \label{eq:Qbr_ST}
          Q_{syn} = \frac{16}{3} \frac{e^2}{c} \left(\frac{eB(x,y,z)}{\mel c}\right)^2 \Thetae^2(x,y,z)~ n_{\rm e}(x,y,z)
      \end{equation}

    where $B$, $n_{\rm e}$, and $\Thetae=kT/ \mel c^2$ are the magnetic field, the number density of electrons, and the dimensionless electron temperature, respectively, at any spatial point $(x,y,z)$. To estimate the electron temperature, we adopt the method described in \citet{2014_Kumar_Chattopadhyay_MNRAS.443.3444K,2025Tripathi_et_al_979_61}. 
    After getting the 3D emission distribution from \eqref{eq:Qbr_ST}, we weight it by the passive tracer distributions ($\Phi$) to account for emission from jet material only. We then perform an integration along the y-direction by summing all the contributions, which yields the synthetic synchrotron surface brightness distribution, denoted as $I(x,z)$ map.
      \begin{equation}
          \label{eq:I_xz}
          I(x,z)=\int \Phi(x,y,z) ~ Q_{syn}(x,y,z)~dy
      \end{equation}
    Here we have assumed a line of sight along the y-axis. Note that the integration is performed over the y-extent of the computational domain, assuming that contributions from outside this region are negligible. The $I(x,z)$ map will help to distinguish between FR I and FR II morphology in our simulations.  
    
\subsection{\label{sec:Jet morphology}Jet morphology in different models} 
%===============================================================    
\subsubsection{\label{sec:ref}Model-REF}
          \begin{figure*}[t]
            \includegraphics[width=\textwidth]{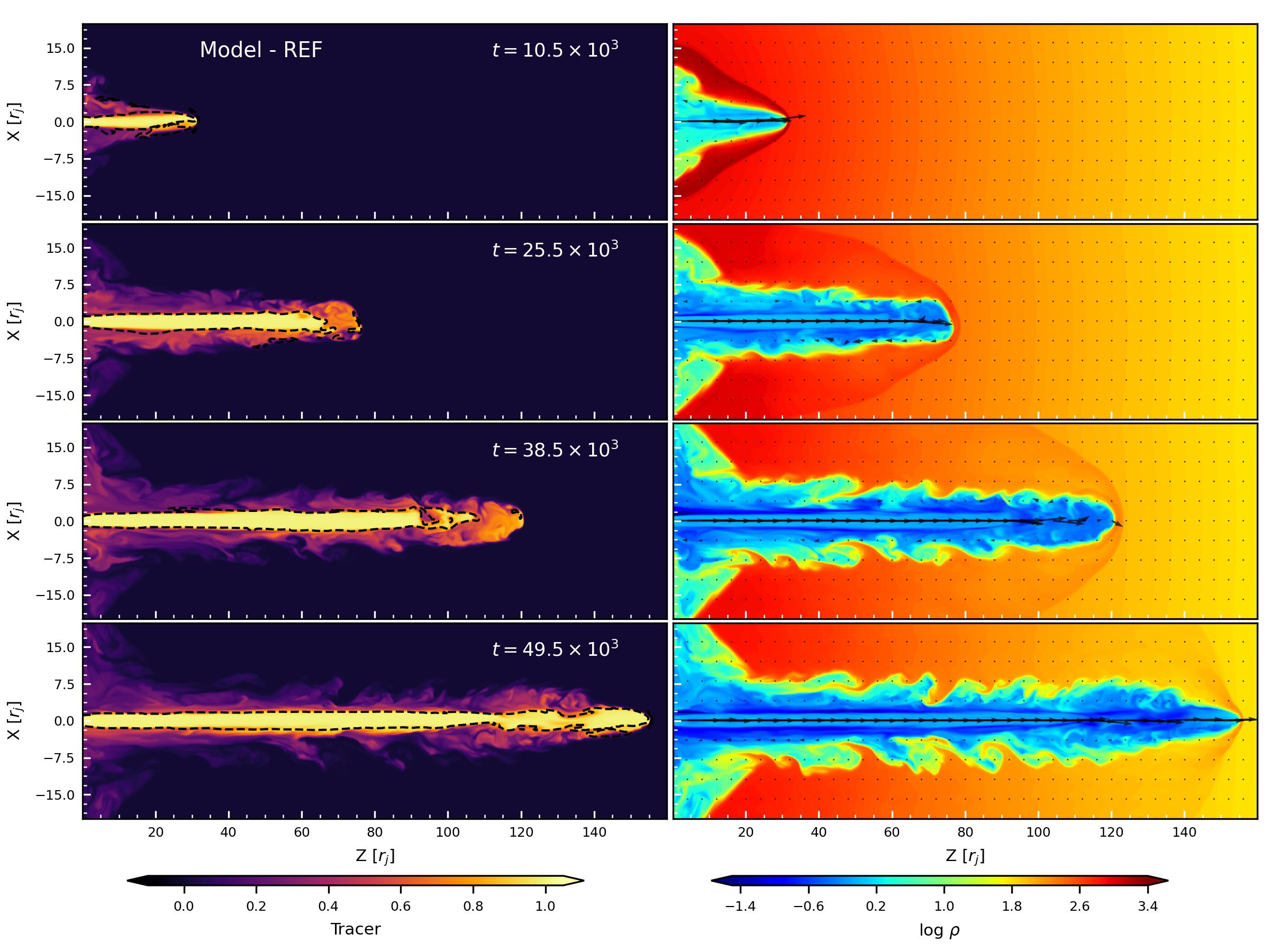}
        \caption{Jet morphology in X-Z plane for Model-REF at various times $t=10.5,25.5,38.5,49.5\times10^3$ in different row panels. The plots are tracer distributions on the left and log density contours on the right, overlaid with velocity vectors (black arrows). The black dashed line in the left panel shows the Mach-1 surface. The length is measured in units of the jet radius $r_j$.}
            \label{fig:REFframes}
          \end{figure*}
    
    Figure~\ref{fig:REFframes} presents the two-dimensional slices of the tracer and logarithmic density distributions in the X-Z plane at various stages of the jet evolution for the reference model (Model-REF). At an early time ($t=10.5\times10^3$), the supersonic jet beam coming from the nozzle interacts with the surrounding ambient medium and generates a forward shock, clearly visible in the density contours (right panel) of Figure~\ref{fig:REFframes}. The left panels also include the Mach-1 surface (black dashed lines), marking the boundary between subsonic and supersonic regions. At initial stages, i.e., $t=10.5\times10^3$, the jet beam remains intact starting from the nozzle up to the jet head, as indicated by the tracer distribution (left panel of Figure~\ref{fig:REFframes}), where the jet tracer value remains close to unity. The jet beam terminates at the forward shock aligned with the location of the Mach-1 surface. As the jet continues to propagate ($t=25.5,\&~38.5\times10^3$), the forward shock gradually weakens due to the inertia imposed by the denser ambient medium. Simultaneously, the jet head also starts to diffuse, as reflected in tracer plots, which show significant mixing of jet and ambient material near the front. However, at $t=49.5\times10^3$ (bottom row of Figure~\ref{fig:REFframes}), the jet beam recovers and extends up to the jethead through the previously diffused region and leads to a stronger forward shock.

\begin{figure}[t]
            \includegraphics[width=\columnwidth]{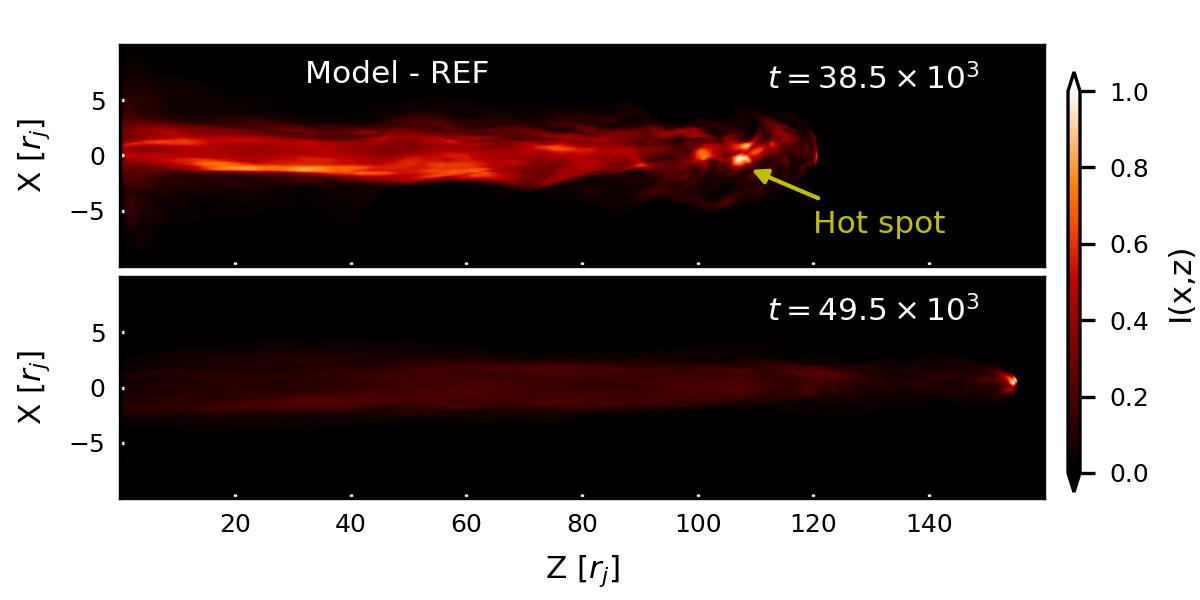}
        \caption{Synthetic synchrotron $I(x,z)$ map for Model-REF at $t=38.5,49.5\times10^3$}
            \label{fig:REFsynchMaps}
          \end{figure}

    It may be noted that the absence or presence of a forward shock works as an indicator of the FR I or FR II morphology in numerical simulations \citep{2022_Massaglia_et_al_A&A}. In FR II jets, the strong forward shock due to supersonic flow results in a hot spot at the jet head where the maximum pressure is located, marking the region of energy dissipation. In contrast, FR I jets lack a prominent forward shock, and the plasma propagates out subsonically \citep{2023_GopalKrishna_JApA}, which is incapable to terminate as a bright hot spot and energy dissipation occurs more gradually along the jet.
    Figure~\ref{fig:REFsynchMaps} shows the synthetic synchrotron emission ($I(x,z)$ map) generated using equation \eqref{eq:I_xz} for model-REF. At $t=38.5\times10^3$, the jet emission is distributed along the entire jet beam and terminates in fainter lobes like a FR I jet. A noticeable hot spot is present at $\sim 108\Rj$. In contrast, at $t=49.5\times10^3$, the jet morphology transitions to an FR II type, where the emission from the jet beam becomes significantly fainter, while a brighter hot spot is seen at $\sim 155\Rj$. This indicates that the jet undergoes morphological changes between FR I and FR II types \citep{1993_DeYoung_ApJ}, and the same source could appear as an FR I or FR II at different times.

%===============================================================  
\subsubsection{\label{sec:hyp}Model-HYP}
          \begin{figure*}[t]
            \includegraphics[width=\textwidth]{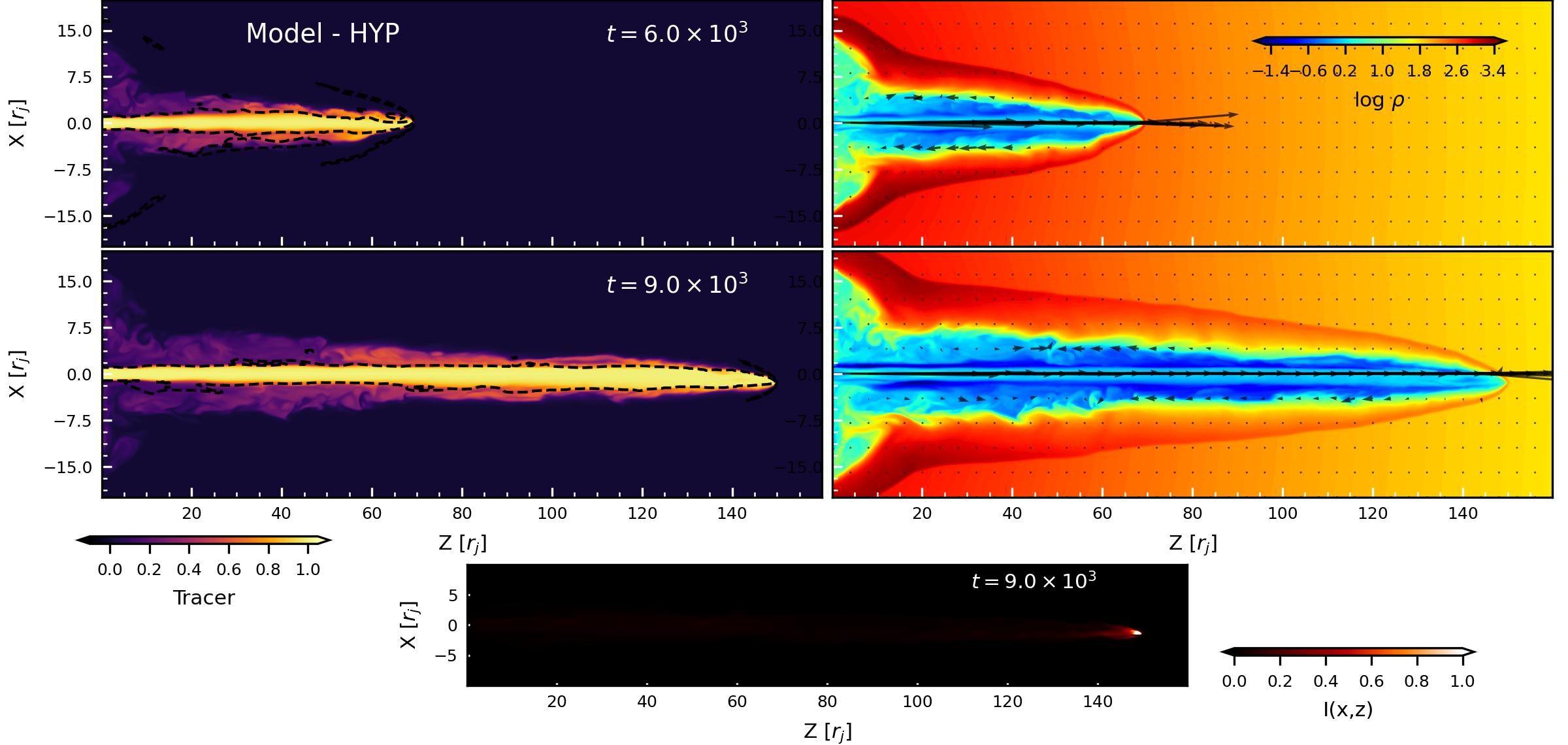} 
        \caption{Top: Jet morphology in X-Z plane for Model-HYP at various times $t=6.0,9.0\times10^3$ in different row panels. The plot variables are the same as in Figure~\ref{fig:REFframes}. Bottom: Synthetic synchrotron $I(x,z)$ map for Model-HYP at $t=9.0\times10^3$}
            \label{fig:HYPframes}
          \end{figure*}

    Figure~\ref{fig:HYPframes} shows the 2D slices of the tracer and logarithmic density distributions in the X-Z plane at two time snaps $t=6\times 10^3$ (top row)
    and $t=9\times 10^3$ of the jet evolution for the Model-HYP. Additionally, a synthetic synchrotron emission ($I(x,z)$ map) is shown at $t=9.0\times10^3$. In this model, all parameters are kept identical to the reference case, except for the injection velocity, which is increased to produce a higher Mach number jet. As evident from the tracer distributions (left panels), the jet beam remains well-collimated and undiffused throughout the simulation. The jet beam propagates up to a distance of $140\Rj$ within a much shorter time $t=9.0\times10^3$ compared to the reference case. Furthermore, a narrower Mach cone of the jet is visible in density contours (right panels), which is a direct consequence of the increased Mach number of the jet flow. 
    A strong forward shock at the jet head gives rise to a prominent hot spot, as seen in the bottom panel of Figure~\ref{fig:HYPframes}. In contrast, the jet beam appears significantly dimmer, characteristic of an FR II morphology, where most of the energy is dissipated at the terminal shock instead of along the jet beam.
    
%===============================================================    
\subsubsection{\label{sec:hot}Model-HOT}
          \begin{figure*}[t]
            \includegraphics[width=\textwidth]{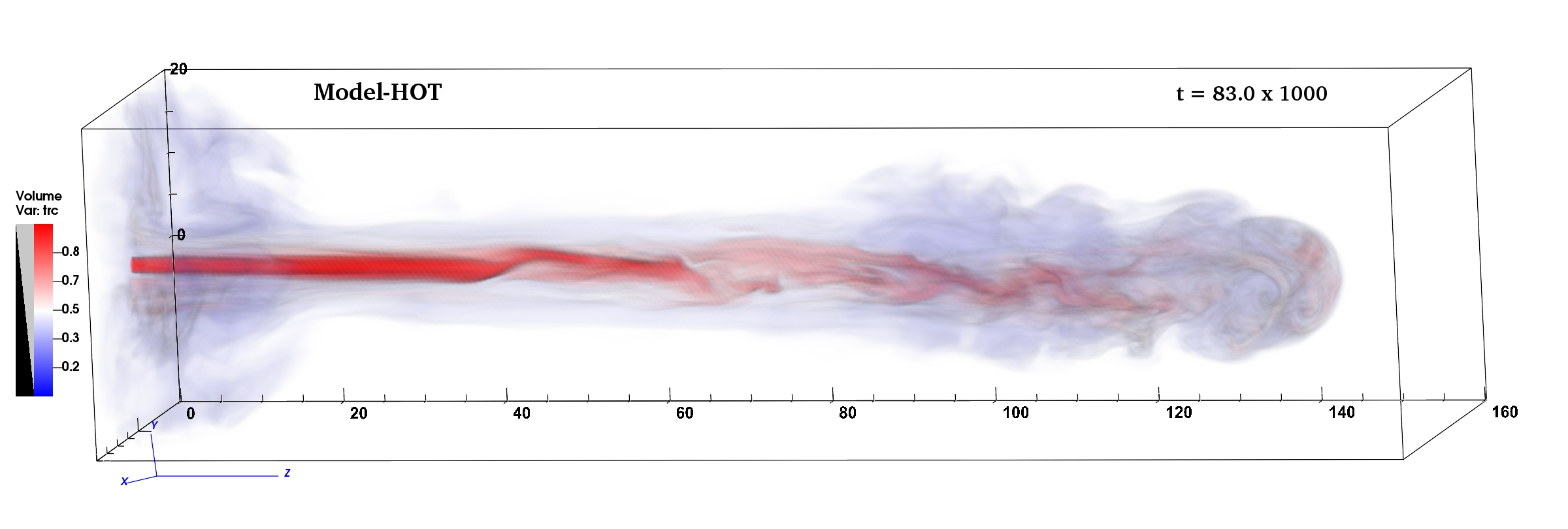}
        \caption{Volume rendering of the tracer distributions for Model-HOT jet at time $t=83.0\times10^3$. (An animation of the figure is available via \href{https://youtu.be/QakcH_pg8U8}{this link.}) The animation shows the propagation of the jet up to $t=85.0\times10^3$, including a camera rotation around the z-axis at $t=41.0\times10^3$ that highlights the development of the kink instability, followed by a spiral zoom at $t=85.0\times10^3$ showing the diffused jet beam.}
            \label{fig:HOTvolume}
          \end{figure*}

    In Model-HOT jet, the core temperature of the ambient medium is increased from $0.167~keV$ to $0.305~keV$ while keeping the jet injection velocity to the same value as the reference case. To maintain the pressure balance of the injected beam with the ambient medium, the temperature of the beam will also be increased. As a result, the jet's Mach number becomes lower. Figure~\ref{fig:HOTvolume} presents a volume rendering of the tracer distributions for the Model-HOT jet at time $t=83.0\times10^3$. We have also included a tracer animation capturing the jet propagation starting from its injection at $t=0.0$. Initially, the supersonic jet beam enters the ambient medium, and due to the onset of the kink instability \citep{1993_Eichler_ApJ,1997_Spruit_et_al,1998_Begelman,2000_Appl_el_al,2006_Giannios_Spruit}, the jet beam begins to develop a wiggling motion (see animation). This wiggling leads to the disruption of the jet beam \citep{2016BrombergTchekhovskoy} and causes the jet head to be diffused (see camera rotation at $t=41.0\times10^3$ in animation). As the jet is continuously injected from the base, the kink instability produces more diffused material at the jet head, to the extent that the jet head is entirely dominated by the diffused material (see spiral zoom at $t=85.0\times10^3$ in animation). 
                  
          \begin{figure*}[t]
            \includegraphics[width=\textwidth]{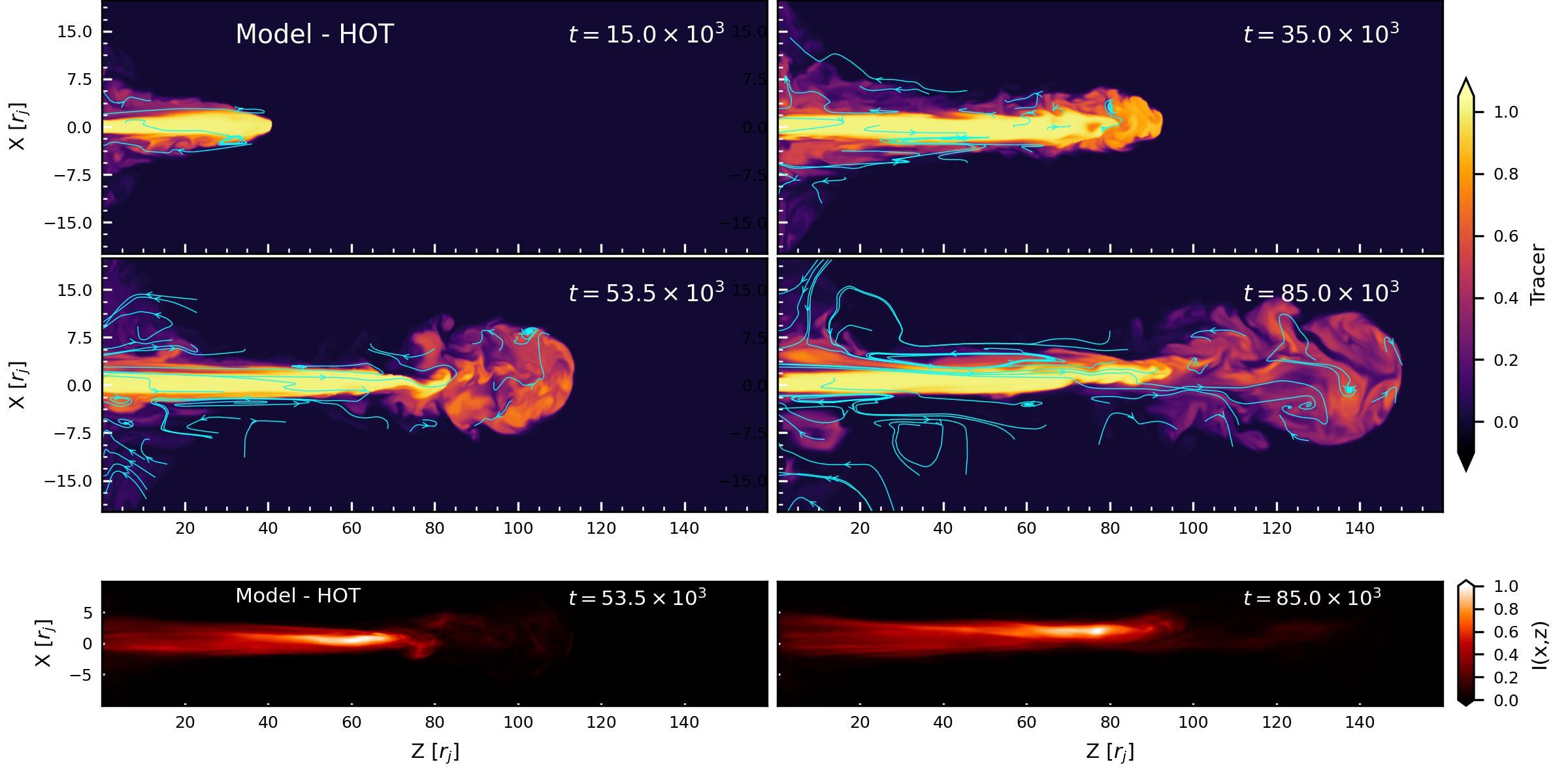}
        \caption{Jet morphology in X-Z plane for Model-HOT at various times $t=15.0,35.0,53.5,85.0\times10^3$ in different panels. The plots in the upper two rows are tracer distributions showing the diffusion of the jet beam overlaid with magnetic field streamlines (in cyan color). The bottom row is the synthetic synchrotron $I(x,z)$ map showing an FR I morphology.}
            \label{fig:HOTframes}
          \end{figure*}
    
    The 2D slices of the snapshots of the tracer distributions in the X-Z plane for the Model-HOT jet are plotted in Figure~\ref{fig:HOTframes} with the $I(x,z)$ map at the last two frames. On the panels of tracer distribution, magnetic field lines are overplotted (cyan curves with arrow heads). At time $t=15.0\times10^3$, the beam remains collimated. By $t=35.0\times10^3$, the onset of diffusion is evident at the head of the jet. At $t=53.5\times10^3$, the jet head is prominently diffused and dispersed into the ambient medium. Additionally, the jet’s collimation has been disrupted due to its weakened Mach disk \citep{1988_GopalKrishnaWiitaNatur.333...49G,2002_Gopal-Krishna_Wiita_NewAR}. Finally, by $t=85.0\times10^3$, we can see the development of a significantly larger and more diffused region propagated up to $150\Rj$. In contrast, the collimated jet beam extends up to a distance of $95\Rj$. Moreover, the propagation of the jet head has become subsonic at these distances. Although only the toroidal magnetic field is injected at the base, jet motion also stretches the magnetic field in the $Z$ direction. The magnetic field has a strong $B_z$ component along the jet beam, while in the diffused cloud, due to disruption of the jet beam, the magnetic field becomes less and less ordered.
    The plasma $\beta$ is very high in the region where the jet beam becomes diffused. The corresponding emission maps show that the jet beam is significantly brighter, with localized regions of enhanced brightness indicating the presence of warm spots. In contrast, the emission from the diffused regions is very low, which means the jet terminates in fainter lobes. In fact, the emission from beyond the distance of $95\Rj$ is nearly negligible compared to that of the earlier beam. This indicates a gradual dissipation of energy along the jet, resulting in an FR I morphology for this model.
    
%===============================================================    
\subsubsection{\label{sec:mag}Model-MAG}
          \begin{figure*}[t]
            \includegraphics[width=\textwidth]{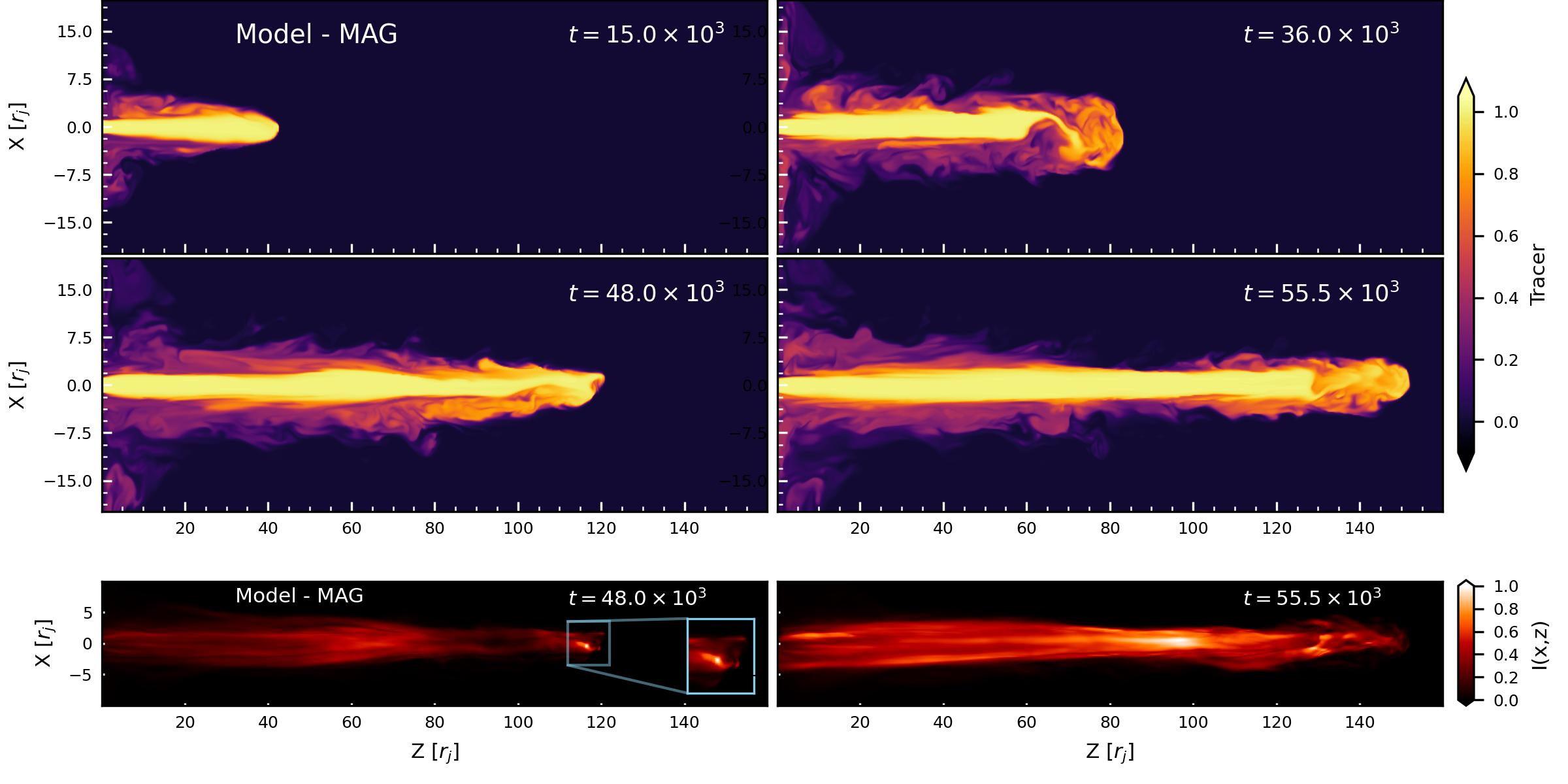}
        \caption{Jet morphology in X-Z plane for Model-MAG at various times $t=15.0,36.0,48.0,55.5\times10^3$ in different panels. The plots in the upper two rows are tracer distributions, and the bottom row is the synthetic synchrotron $I(x,z)$ map.}
            \label{fig:MAGframes}
          \end{figure*}
          
    Figure~\ref{fig:MAGframes} shows the 2D slices of the tracer distributions in the X-Z plane at different stages of the jet evolution with the $I(x,z)$ map at the last two frames for the Model-MAG. In this model, the magnetization of the jet is increased while all other parameters remain the same as in the reference case. The strength of the magnetization is quantified by the plasma-beta ($\beta$) parameter, which represents the ratio of gas pressure to magnetic pressure. A lower $\beta$ value corresponds to a stronger magnetic field in the jet and vice versa. To increase magnetization, the beta value is reduced from 10 to 4. As shown by the tracer distributions in Figure~\ref{fig:MAGframes}, the jet beam initially remains collimated and undiffused at $t=15.0\times10^3$. By $t=36.0\times10^3$, signs of diffusion become apparent; however, by $t=48.0\times10^3$, the jet beam penetrates the previously diffused region and reestablishes its collimated structure. This behavior is mirrored in the corresponding synchrotron emission map, which displays a fainter jet beam along its length and a bright warm spot \citep{2019_Massaglia_et_al_A&A} at $\sim 120\Rj$. Note that this warm spot is not due to the terminal shock because there is a fainter emission ahead of this bright region (see inset), and that's why it differs from the classical hot spots present in powerful FR II radio jets. In later stages at $t=55.5\times10^3$, one can again see the development of a diffused head in the tracer distribution. The $I(x,z)$ map at this time frame displays the sign of an FR I morphology, with localized bright patches.
    Higher magnetization enhances the jet's collimation in the first place, as backflow within the jet amplifies the magnetic field strength near its base. This magnetic field, predominantly azimuthal, supports the collimation. However, in later stages, we anticipate the growth of strong non-axisymmetric current-driven instabilities, which ultimately destabilize the jet \citep{1993_Eichler_ApJ} and cause a sudden disruption. The jet will release the bulk of its energy, giving rise to bright warm spots, beyond which the flow continues to propagate, giving rise to fainter regions as discussed by \citet{2019_Massaglia_et_al_A&A}.

%=============================================================================
\subsection{\label{sec:composition effects}Can plasma composition change the morphology?}
%===============================================================
          \begin{figure*}[t]
            \includegraphics[width=\textwidth]{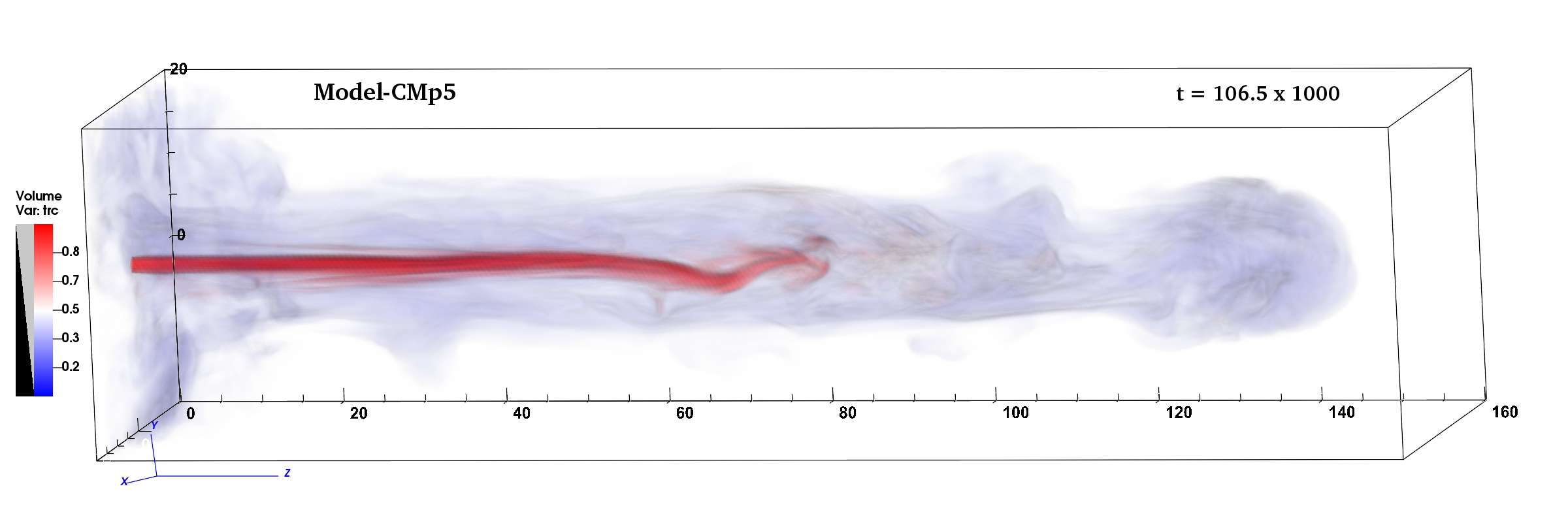}
        \caption{Volume rendering of the tracer distributions for Model-CMp5 jet at time $t=106.5\times10^3$. (An animation of the figure is available via \href{https://youtu.be/PRVU6AeWMmo}{this link.}) The animation shows the propagation of the jet up to $t=108.5\times10^3$, including a camera rotation around the z-axis at $t=32.5\times10^3$, followed by a camera rotation at $t=108.5\times10^3$ showing the diffused jet beam.}
            \label{fig:CMp5volume}
          \end{figure*}

    Until now, all models have considered jets composed purely of electron-proton plasma. Now, we extend our analysis to include models that incorporate multi-species plasmas, particularly positrons, into the jet composition. We tune our composition parameter ($\xi$) defined as the ratio of the proton number density to the electron number density. The $\xi$ parameter directly affects the internal energy of the flow through CR EoS (see eqn~\eqref{eq:CR EoS}), thereby affecting the thermodynamic properties of the jet.
%===============================================================
\subsubsection{\label{sec:cmp5}Model-CMp5}
    We first examine the Model-CMp5 jet configuration, in which the composition parameter ($\xi$) is fixed at 0.5. This corresponds to a plasma consisting of equal proportions of protons and positrons. All other injection parameters of the jet are the same as in the reference model. At a given physical temperature, the sound speed for the lepton-rich flow (i.e., positrons) will be higher. This is because the sound speed depends on the inverse of the square root of the mass. Consequently, for the same jet injection velocity, the Mach number in this multi-species plasma case is reduced relative to the reference proton-electron case. This reduction in Mach number causes weakening in the Mach disk, which has a significant influence in determining the jet’s collimation \citep{1988_GopalKrishnaWiitaNatur.333...49G} and morphology, which we will see further.  
    
    Figure~\ref{fig:CMp5volume} presents a volume rendering of the tracer distributions for the Model-CMp5 jet at time $t=106.5\times10^3$. (An animation of the jet propagation is also available.) The jet beam coming from the nozzle initially retains its well-collimated beam structure until the kink instability starts to develop. This instability allows the jet beam to go off-axis and exhibit a noticeable wiggling motion (see animation). The jet head starts to diffuse as apparent in camera rotation at $t=32.5\times10^3$ in the animation. As the jet moves forward, the wiggling motion disrupts the jet beam when it tries to extend further (see animation). Consequently, more diffused material accumulates at the jet head, which is no longer a collimated structure but a broad distribution of tracer material. This diffused material continues to propagate, eventually reaching beyond $140\Rj$, well ahead of the beam itself. Over time, the jet head becomes fully dominated by this diffused material, with the original beam left far behind (see Figure~\ref{fig:CMp5volume} and the camera rotation at $t=108.5\times10^3$ in the animation).
    
%===============================================================
          \begin{figure*}[t]
            \includegraphics[width=\textwidth]{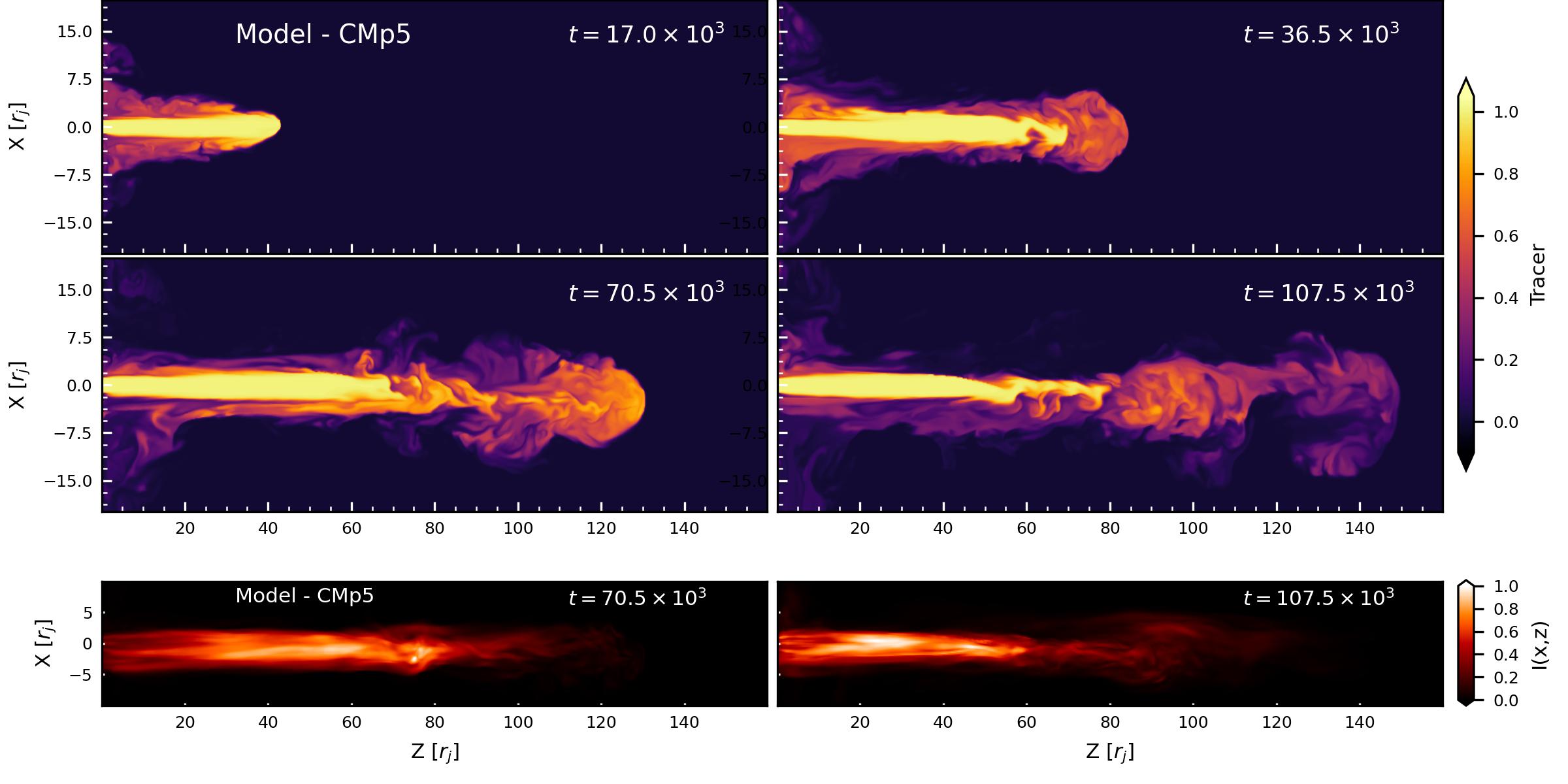}
        \caption{Jet morphology in X-Z plane for Model-CMp5 at various times $t=17.0,36.5,70.5,107.5\times10^3$ in different panels. The plots in the upper two rows are tracer distributions and the bottom row is the synthetic synchrotron $I(x,z)$ map.}
            \label{fig:CMp5frames}
          \end{figure*}

    The 2D slices of the tracer distributions in the X-Z plane for the Model-CMp5 jet are plotted in Figure~\ref{fig:CMp5frames} at different stages of its evolution with the $I(x,z)$ map at the last two frames. Initially, the jet beam appears as a well-collimated structure at $t=17.0\times10^3$, with no evidence of diffusion. However, the signs of diffusion are evident at the head of the jet at $t=36.5\times10^3$, where the material at the front of the jet spreads laterally. As the jet evolves further, we can see that the jet head becomes significantly more diffuse and broadened in tracer panels shown at $t=70.5,107.5\times10^3$ as discussed previously. 
    The corresponding emission maps also show minimal emission from diffused regions, making them fainter lobes, while the beam itself contains relatively brighter, localized patches. This indicates a persistent FR I morphology for this model, where the jet loses much of its energy gradually along the beam rather than delivering it efficiently to the jet head.

%===============================================================
\subsubsection{\label{sec:cmp2}Model-CMp2}
          \begin{figure*}[t]
            \includegraphics[width=\textwidth]{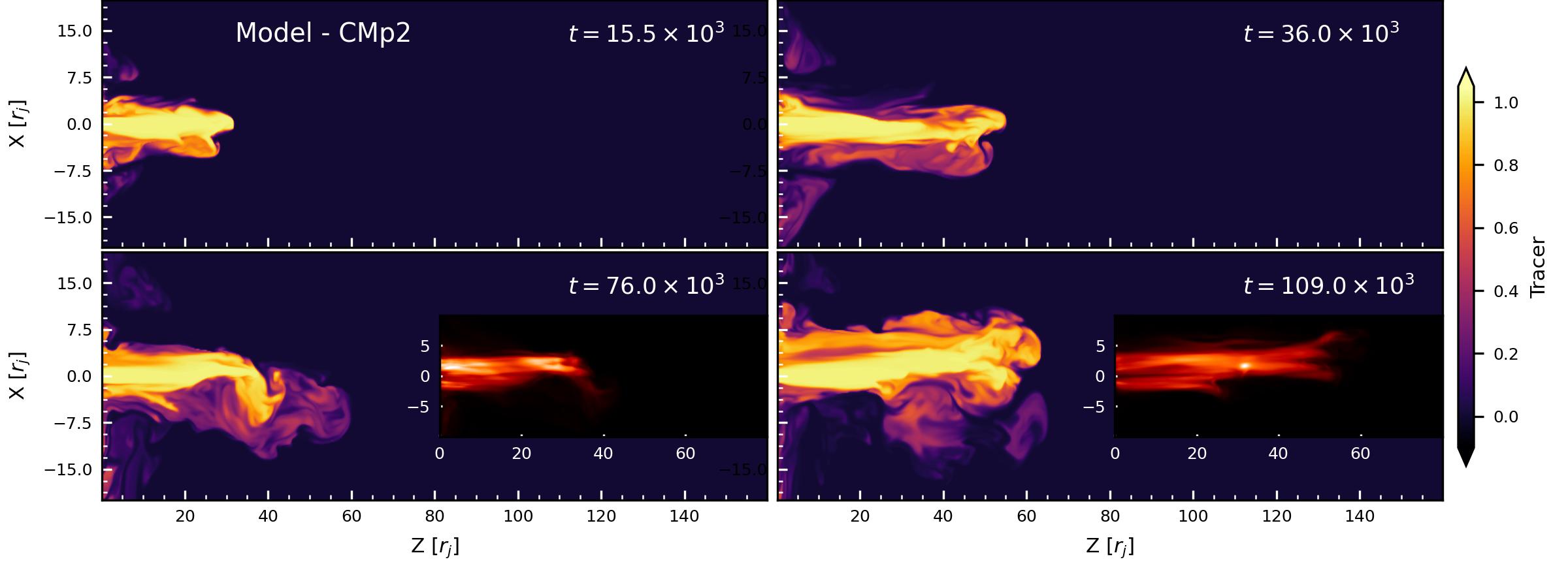}
        \caption{Jet morphology in X-Z plane for Model-CMp2 at various times $t=15.5,36.0,76.0,109.0\times10^3$ in different panels. The plots are the tracer distributions, and the inset shows the corresponding synthetic synchrotron $I(x,z)$ map.}
            \label{fig:CMp2frames}
          \end{figure*}

    Figure~\ref{fig:CMp2frames} presents 2D slices of the tracer distributions in the X–Z plane for the Model-CMp2 jet at various stages of its evolution. Model CMp2 corresponds to $\xi=0.2$ or the proton number density is 20\% of the electron number density, and the rest is positrons. Insets in the two lower panels show the corresponding $I(x,z)$ emission maps. Even as early as $t=15.5\times10^3$, it is evident that the jet beam struggles to maintain a straight course along $z$. At $t=36.0\times10^3$, the jet beam tries to find a new path to expand. At $t=76.0\times10^3$, the kink instability becomes more pronounced, as seen by its bent structure in the tracer map. By $t=109.0\times10^3$, the jet advance stalls. It is clear that the jet is unable to propagate beyond a distance of $60\Rj$, indicating it has reached a stagnation point where the jet's energy is no longer sufficient to continue driving the head forward. The corresponding emission maps display an FR I-type morphology, with some local bright patches and the absence of a forward shock. A hotspot is also visible within the jet beam around $30\Rj$ in the emission map at $t=109.0\times10^3$. 

%=============================================================================
\subsection{\label{sec:head evolutions}Time evolution of the jet heads}

          \begin{figure*}[t]
            \includegraphics[width=0.49\textwidth]{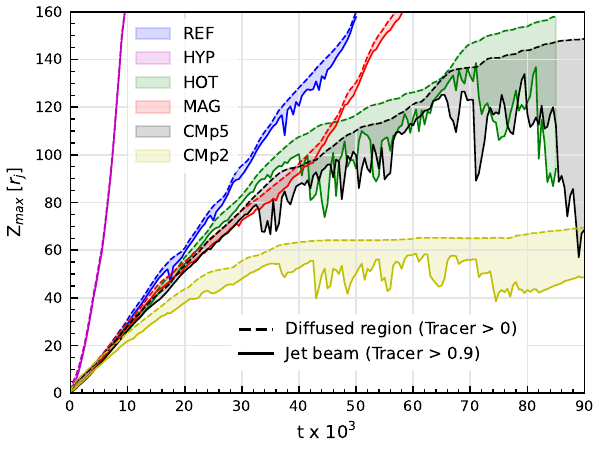}
            \includegraphics[width=0.49\textwidth]{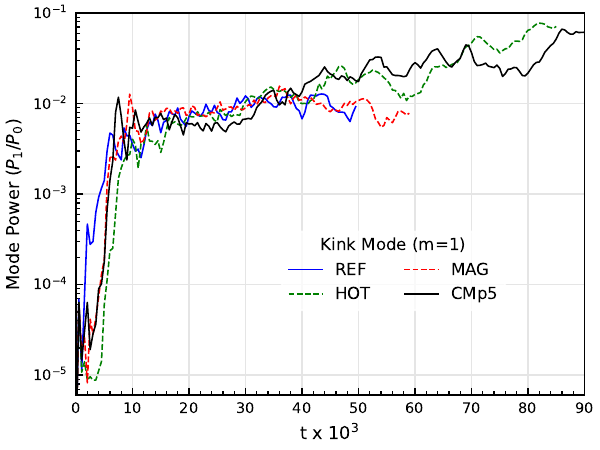}
        \caption{Left: evolution of the jet head over time for all models is shown in different colors. The dashed and solid lines represent the maximum axially propagated distance of the diffused jet material and the jet beam, respectively. The color-shaded region shows the length of the diffused region. Right: temporal evolution of the kink (m=1) mode power for four of the models.}
            \label{fig:JetHeadEvolution}
          \end{figure*}     
    To gain a better understanding of the temporal behavior of various jet models discussed in this study, we perform an analysis of jet head diffusion over time. The left panel of figure~\ref{fig:JetHeadEvolution} compares the temporal evolution of the jet head of all the models. Each of the six jet models is represented by a distinct color. To quantify the extent of jet propagation, we identify two axial positions along the Z-direction. First, we identify the furthest point where the tracer value exceeds 0, which marks the leading edge of the jet material. Second, we identify the furthest position where the tracer value remains above 0.9, which we interpret as the jet core regions, minimally affected by the surrounding medium. Both of these positions are plotted over time using dashed and solid lines, respectively, to visualize the axial propagation of the jet. Additionally, the region between these two curves is color-shaded to visualize the spatial length of the diffused region at each time step. 
    In the right panel of figure~\ref{fig:JetHeadEvolution}, we show the growth rate of the kink mode (m=1) power calculated for four models shown in different colors. The growth of the non-axisymmetric m=1 mode indicates the displacement of the jet from its axis, resulting in the deformation of the jet spine. The mode power is calculated from the Fourier decomposition in the azimuthal ($\phi$) direction, and averaged in the r and z-directions. It may be noted that we checked, the m=1 mode dominates over other non-axisymmetric modes, but has not been presented to avoid cluttering Figure~\ref{fig:JetHeadEvolution} (right panel).
    
    As shown in the left panel of Figure~\ref{fig:JetHeadEvolution}, all jet models show an FR II morphology during the initial phases, where there is no diffusion of the jet head. The jet head of the Model-REF (blue color) undergoes slight diffusion during its propagation, indicated by the increased separation of the two curves (dashed \& solid), and they converge again (marked by the extent of the color-shaded region). During the phase of diffusion, the jet temporarily transits to FR I morphology. However, it eventually transits back to FR II morphology as the jet advances farther out. By the end of the simulation, the jet beam covers a length of $160\Rj$ in $t \sim 50.0\times10^3$. In the right panel, the kink mode power (blue) also rises and eventually decays before leaving the computational domain. In contrast, Model-HYP (magenta) shows no diffusion throughout its evolution. It covers the same length of $160\Rj$ in a significantly shorter time $\sim 10.0\times10^3$ due to higher jet velocity and consistently maintains its FR II morphology. In Model-HOT (green), a higher thermal resistance from the ambient medium is offered to the jet, which slows its propagation. 
    The kink feature shown in Figure~\ref{fig:HOTvolume} at $t=41.0\times10^3$ (camera rotation in animation), where the jet stream bends from the axis, is also indicated by the continuous growth of the kink mode power (right panel, Figure~\ref{fig:JetHeadEvolution}). 
    The jet head diffuses significantly, indicated by the green shaded region in left panel. As the jet beam becomes completely disrupted at later times, the jet head becomes dominated by diffused material (see also Figure~\ref{fig:HOTvolume}~\&~\ref{fig:HOTframes}). This model retains its FR I morphology throughout. In Model-MAG (red), the propagation of the jet beam starts slightly slower compared to Model-REF. Kink instability sets in at $t\sim 10\times 10^3$ and starts to disrupt the jet head; however, at later times $t>40\times 10^3$, the kink mode (red curve, right panel) gets dynamically suppressed, and the jet head is less dissipative; therefore, this model also switches between FR I and II morphology. 
    In the fifth Model-CMp5 (black), the lepton-rich jet flow also exhibits slow propagation, and the diffusion in the jet head gradually increases with time due to amplification of the kink mode instability, eventually leading to a complete disruption of the jet beam beyond a certain extent. This is also evident by the growth in kink mode power. As a consequence, the jet exhibits a persistent FR I morphology. In the last Model-CMp2 (yellow), the jet's forward propagation significantly diminishes and stagnates at $t > 40.0\times10^3$. Beyond this time, the jet loses nearly all of its energy within a short distance, and therefore, this model also exhibits an FR I morphology.

          \begin{figure}[h]
            \includegraphics[width=\columnwidth]{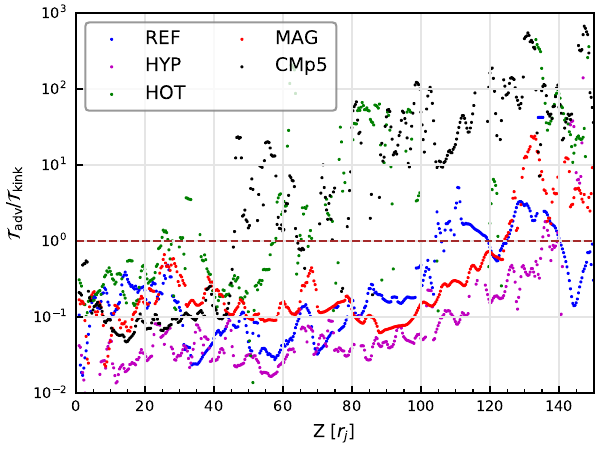}
            \caption{Ratio of the advection timescale $(z/v_z)$ to kink growth timescale along the axis of the jet for five models in different colors.}
            \label{fig:kinkGrowthTimescale}
          \end{figure}

    If we compare the kink mode growth timescale with the advection timescale in the jet, some interesting facts emerge. Linear stability analysis shows that the maximum growth rate of the kink mode is $\lambda_{\rm max}= 0.133v_A/P_0$ \citep{2000_Appl_el_al,2013Bodo_MNRAS,2019Bromberg_et_alApJ}. Here, the expression of $\lambda_{\rm max}$ has been obtained assuming negligible thermal pressure, and $v_A$ is the Alfven velocity, and $P_0=rB_z/B_{\phi}$ is the pitch of the magnetic field at the axis. Therefore, the e-folding growth time of the most unstable kink mode will be $\mathcal{T}_{\rm kink}= 1/ \lambda_{\rm max}$.
    A jet would be disrupted by the kink mode if $\mathcal{T}_{\rm adv}/\mathcal{T}_{\rm kink} \geq 1$; otherwise, the mode will be advected away before it grows \citep{2016BrombergTchekhovskoy, 2016_Tchekhovskoy_Bromberg_MNRAS}.
    In figure~\ref{fig:kinkGrowthTimescale}, we have plotted the $\mathcal{T}_{\rm adv}/\mathcal{T}_{\rm kink}$ along the axis of the jet for five models at times just before the jet leaves the domain. Model REF (blue) is slightly kink unstable ($\mathcal{T}_{\rm adv}/\mathcal{T}_{\rm kink} \geq 1$) at the head of the jet, where the jet speed ($v_z$) is lower than in the jet spine. In model HOT (green), the temperature of the jet at the injection point is about twice the value in REF, and $\mathcal{T}_{\rm adv}/\mathcal{T}_{\rm kink} \geq 1$ almost the entire length of the jet. Due to higher temperature, the jet beam would tend to expand in the lateral direction, which would suppress the development of $B_z$ and enhance the kink growth rate (see the expression of $\lambda_{\rm max}$ above). For model HYP (magenta), higher axial velocity at the injection nozzle reduces the advection timescale throughout the whole jet beam, such that it remains less than the kink growth timescale. Higher jet speed implies two things: one, $\mathcal{T}_{\rm adv}<\mathcal{T}_{\rm kink}$ so the instability will not get enough time to develop before being advected away, and two, higher $v_z$ will stretch the magnetic field and generate $B_z$, which would suppress the kink growth rate. It may be noted here that for the same Mach number jet, the one with lower injected speed would have a larger advection timescale and would be susceptible to various instabilities.
    By increasing the injected $B_\phi$ in model MAG, but keeping the other parameters of model REF, model MAG showed increased kink instability (red).
    Jets composed of a mixture of protons, electrons, and positrons (Model CMp5), i.e., for $\xi < 1$, have proportionately less protons (baryons) than leptons. And flows at the same temperature but less protons are relatively thermally hotter than flows with more protons. This has been explained in a series of papers \citep[for e.g.][]{2009Chattopadhyay&RyuApJ,2011_Chattopadhyay_Chakrabarti_IJMPD..20.1597C}. So for models $\xi<1$ (black), reduction of the proton proportion made the flow relatively much hotter than the electron-proton jet. And as explained with the HOT model, the hot beam laterally expands, which inhibits the development of $B_z$ and therefore contributes to the kink growth rate.

%%%%%%%%%%%%%%%%%%%%%%%%%%%%%%%%%%%%%%%%%%%%%%%%%%
\section{\label{sec:Summary and Discussions}Summary and Discussions}
    We have carried out numerical studies of low-power FR I jets at the kpc scales using large-scale 3D MHD simulations. We employed a finite volume, second-order Godunov-type scheme in our simulations. The thermodynamics of the gas is governed by the variable $\Gamma$ CR EoS, which can also account for the composition effects, such as an arbitrary mixture of electrons, protons, and positrons. Moreover, we also generated the posterior synchrotron emission map from our simulation results to distinguish between FR I and II morphology. We simulated a total of six cases, the results of which are summarized below:
    \begin{enumerate}
        \item The reference jet model showed transient morphological changes between FR I and FR II types in its later stages of evolution, suggesting that the observation epoch can also play a role in distinguishing between the FR I and II types. Additionally, some bright spots were also observed in emission maps.
        \item The second case is a highly supersonic ($\mathcal{M}_{\rm j}=10$ ) jet model, having the highest jet kinetic power among the models investigated. It evolved into FR II morphology with a brighter hot spot at its head because of a strong termination shock.
        \item The third model (HOT), where we injected the jet in a hotter ambient medium, evolved into an FR I morphology only, because of a weakened Mach disk. The jet beam is destabilized at the head because of the onset of kink instability, and the wiggling beam dissipates, producing a diffused region. The synchrotron map showed an FRI morphology.
        \item The fourth model (MAG), where the magnetization of the jet was increased, also shows a transient nature in morphology. It is expected that the increased toroidal magnetic field will result in enhanced collimation, but by the same token, it will also give rise to strong non-axisymmetric kink instabilities, causing disruption of the jet beam.
        \item Growth of various instabilities in jets with higher injection speed is minimized, to the extent that, if two jets are launched with the same Mach number, then the jet with low injection speed might be susceptible to instabilities like kink modes.
        \item Next, we explored the effect of plasma composition in Model-CMp5 and CMp2. The presence of lepton-rich flow resulted in an increase in the thermal energy of the jet. Consequently, the jet injection Mach number was also decreased, causing a weakened Mach disk. As a result, both of these models evolved into FR I morphology. So, a higher leptonic proportion may lead to a greater possibility of a disrupted jet beam and consequently FRI-type morphology. It is interesting that with the increase in lepton proportion, the jet seems to stall at a shorter distance from the base. While the presence of heavier particles like protons in the gas, the length scales are much larger, even if the flow is FR I type. 
    \end{enumerate}

    We further showed that jets in which the kink mode grows on a timescale shorter than the advection timescale are more susceptible to disruption, leading to an FR I type morphology.
    In this work, we have generated the emission map assuming that the radiation emitted is not disturbed anywhere on its way to reaching the observer. It gives a crude estimate of the jet morphology; however, the use of slightly better synchrotron emission modeling \citep{2018VaidyaApJ,2019_vanderWesthuizen,uvs24} will give a better picture.

\section*{Acknowledgment}
 The authors acknowledge the anonymous referee for fruitful suggestions, which significantly improved the quality of the paper.
The authors gratefully acknowledge the use of computational resources provided by ARIES’s \texttt{Surya} and IUCAA’s \texttt{Pegasus} HPC clusters for conducting the 3D simulations presented in this work. RC thanks ANRF for a SURE grant (SUR/2022/001503) and IUCAA for their hospitality and usage of their facilities during his stay at different times as part of the university associateship program. The underlying simulation data will be made available upon reasonable request to the corresponding author.
%\begin{acknowledgments}
%
%\end{acknowledgments}


%\vspace{5mm}
%\facilities{HST(STIS), Swift(XRT and UVOT), AAVSO, CTIO:1.3m,CTIO:1.5m,CXO}


%\software{astropy \citep{2013A&A...558A..33A,2018AJ....156..123A},  Cloudy \citep{2013RMxAA..49..137F}, Source Extractor \citep{1996A&AS..117..393B}, Scipy (Jones et al. 2001)}
\software{\texttt{Python} \citep{2007_Oliphant_Python}, \texttt{NumPy} \citep{2020_harris_NumPy}, \texttt{Matplotlib} \citep{2007_Hunter_Matplotlib}, \texttt{VisIt} \citep{2012_Childs_VisIt}}



%%%%%%%%%%%%%%%%%%%%%%%%%%%%%%%%%%%%%%%%%%%%%%%%%%
%\appendix

\bibliography{jet_references}{}
\bibliographystyle{aasjournal}



\end{document}